\def\unlock{\catcode`@=11} 
\def\lock{\catcode`@=12} 
\unlock
%
%
%
%
%

\font\fourteenrm=cmr10 scaled\magstep2
\font\twelverm=cmr10 scaled\magstep1
\font\ninerm=cmr9          \font\sixrm=cmr6

\font\fourteenbf=cmbx10 scaled\magstep2
\font\twelvebf=cmbx10 scaled\magstep1

\font\seventeeni=cmmi10 scaled\magstep3     \skewchar\seventeeni='177
\font\fourteeni=cmmi10 scaled\magstep2      \skewchar\fourteeni='177
\font\twelvei=cmmi10 scaled\magstep1        \skewchar\twelvei='177
\font\ninei=cmmi9                           \skewchar\ninei='177
\font\sixi=cmmi6                            \skewchar\sixi='177
\font\seventeensy=cmsy10 scaled\magstep3    \skewchar\seventeensy='60
\font\fourteensy=cmsy10 scaled\magstep2     \skewchar\fourteensy='60
\font\twelvesy=cmsy10 scaled\magstep1       \skewchar\twelvesy='60
\font\ninesy=cmsy9                          \skewchar\ninesy='60
\font\sixsy=cmsy6                           \skewchar\sixsy='60

\font\fourteenex=cmex10 scaled\magstep2
\font\twelveex=cmex10 scaled\magstep1

\font\fourteensl=cmsl10 scaled\magstep2
\font\twelvesl=cmsl10 scaled\magstep1

\font\fourteenit=cmti10 scaled\magstep2
\font\twelveit=cmti10 scaled\magstep1
\font\twelvett=cmtt10 scaled\magstep1
\font\twelvecp=cmcsc10 scaled\magstep1
\font\tencp=cmcsc10
\newfam\cpfam
%
%
\newcount\f@ntkey            \f@ntkey=0
\def\samef@nt{\relax \ifcase\f@ntkey \rm \or\oldstyle \or\or
         \or\it \or\sl \or\bf \or\tt \or\caps \fi }
\def\fourteenpoint{\relax
    \textfont0=\fourteenrm          \scriptfont0=\tenrm
    \scriptscriptfont0=\sevenrm
     \def\rm{\fam0 \fourteenrm \f@ntkey=0 }\relax
    \textfont1=\fourteeni           \scriptfont1=\teni
    \scriptscriptfont1=\seveni
     \def\oldstyle{\fam1 \fourteeni\f@ntkey=1 }\relax
    \textfont2=\fourteensy          \scriptfont2=\tensy
    \scriptscriptfont2=\sevensy
    \textfont3=\fourteenex     \scriptfont3=\fourteenex
    \scriptscriptfont3=\fourteenex
    \def\it{\fam\itfam \fourteenit\f@ntkey=4 }\textfont\itfam=\fourteenit
    \def\sl{\fam\slfam \fourteensl\f@ntkey=5 }\textfont\slfam=\fourteensl
    \scriptfont\slfam=\tensl
    \def\bf{\fam\bffam \fourteenbf\f@ntkey=6 }\textfont\bffam=\fourteenbf
    \scriptfont\bffam=\tenbf     \scriptscriptfont\bffam=\sevenbf
    \def\tt{\fam\ttfam \twelvett \f@ntkey=7 }\textfont\ttfam=\twelvett
    \h@big=11.9\p@{} \h@Big=16.1\p@{} \h@bigg=20.3\p@{} \h@Bigg=24.5\p@{}
    \def\caps{\fam\cpfam \twelvecp \f@ntkey=8 }\textfont\cpfam=\twelvecp
    \setbox\strutbox=\hbox{\vrule height 12pt depth 5pt width\z@}
    \samef@nt}
\def\twelvepoint{\relax
    \textfont0=\twelverm          \scriptfont0=\ninerm
    \scriptscriptfont0=\sixrm
     \def\rm{\fam0 \twelverm \f@ntkey=0 }\relax
    \textfont1=\twelvei           \scriptfont1=\ninei
    \scriptscriptfont1=\sixi
     \def\oldstyle{\fam1 \twelvei\f@ntkey=1 }\relax
    \textfont2=\twelvesy          \scriptfont2=\ninesy
    \scriptscriptfont2=\sixsy
    \textfont3=\twelveex          \scriptfont3=\twelveex
    \scriptscriptfont3=\twelveex
    \def\it{\fam\itfam \twelveit \f@ntkey=4 }\textfont\itfam=\twelveit
    \def\sl{\fam\slfam \twelvesl \f@ntkey=5 }\textfont\slfam=\twelvesl
    \scriptfont\slfam=\ninerm
    \def\bf{\fam\bffam \twelvebf \f@ntkey=6 }\textfont\bffam=\twelvebf
    \scriptfont\bffam=\ninerm     \scriptscriptfont\bffam=\sixrm
    \def\tt{\fam\ttfam \twelvett \f@ntkey=7 }\textfont\ttfam=\twelvett
    \h@big=10.2\p@{}
    \h@Big=13.8\p@{}
    \h@bigg=17.4\p@{}
    \h@Bigg=21.0\p@{}
    \def\caps{\fam\cpfam \twelvecp \f@ntkey=8 }\textfont\cpfam=\twelvecp
    \setbox\strutbox=\hbox{\vrule height 10pt depth 4pt width\z@}
    \samef@nt}
\def\tenpoint{\relax
    \textfont0=\tenrm          \scriptfont0=\sevenrm
    \scriptscriptfont0=\fiverm
    \def\rm{\fam0 \tenrm \f@ntkey=0 }\relax
    \textfont1=\teni           \scriptfont1=\seveni
    \scriptscriptfont1=\fivei
    \def\oldstyle{\fam1 \teni \f@ntkey=1 }\relax
    \textfont2=\tensy          \scriptfont2=\sevensy
    \scriptscriptfont2=\fivesy
    \textfont3=\tenex          \scriptfont3=\tenex
    \scriptscriptfont3=\tenex
    \def\it{\fam\itfam \tenit \f@ntkey=4 }\textfont\itfam=\tenit
    \def\sl{\fam\slfam \tensl \f@ntkey=5 }\textfont\slfam=\tensl
    \def\bf{\fam\bffam \tenbf \f@ntkey=6 }\textfont\bffam=\tenbf
    \scriptfont\bffam=\sevenbf     \scriptscriptfont\bffam=\fivebf
    \def\tt{\fam\ttfam \tentt \f@ntkey=7 }\textfont\ttfam=\tentt
    \def\caps{\fam\cpfam \tencp \f@ntkey=8 }\textfont\cpfam=\tencp
    \setbox\strutbox=\hbox{\vrule height 8.5pt depth 3.5pt width\z@}
    \samef@nt}
%
%
%
%
\newdimen\h@big  \h@big=8.5\p@
\newdimen\h@Big  \h@Big=11.5\p@
\newdimen\h@bigg  \h@bigg=14.5\p@
\newdimen\h@Bigg  \h@Bigg=17.5\p@
\def\big#1{{\hbox{$\left#1\vbox to\h@big{}\right.\n@space$}}}
\def\Big#1{{\hbox{$\left#1\vbox to\h@Big{}\right.\n@space$}}}
\def\bigg#1{{\hbox{$\left#1\vbox to\h@bigg{}\right.\n@space$}}}
\def\Bigg#1{{\hbox{$\left#1\vbox to\h@Bigg{}\right.\n@space$}}}
%
%
%
\normalbaselineskip = 20pt plus 0.2pt minus 0.1pt
\normallineskip = 1.5pt plus 0.1pt minus 0.1pt
\normallineskiplimit = 1.5pt
\newskip\normaldisplayskip
\normaldisplayskip = 18pt plus 4pt minus 8pt
\newskip\normaldispshortskip
\normaldispshortskip = 5pt plus 4pt
\newskip\normalparskip
\normalparskip = 6pt plus 2pt minus 1pt
\newskip\skipregister
\skipregister = 5pt plus 2pt minus 1.5pt
\newif\ifsingl@    \newif\ifdoubl@
\newif\iftwelv@    \twelv@true
\def\singlespace{\singl@true\doubl@false\spaces@t}
\def\doublespace{\singl@false\doubl@true\spaces@t}
\def\normalspace{\singl@false\doubl@false\spaces@t}
\def\Tenpoint{\tenpoint\twelv@false\spaces@t}
\def\Twelvepoint{\twelvepoint\twelv@true\spaces@t}
\def\spaces@t{\relax%
 \iftwelv@ \ifsingl@\subspaces@t3:4;\else\subspaces@t1:1;\fi%
 \else \ifsingl@\subspaces@t3:5;\else\subspaces@t4:5;\fi \fi%
 \ifdoubl@ \multiply\baselineskip by 5%
 \divide\baselineskip by 4 \fi \unskip}
\def\subspaces@t#1:#2;{%
      \baselineskip = \normalbaselineskip%
      \multiply\baselineskip by #1 \divide\baselineskip by #2%
      \lineskip = \normallineskip%
      \multiply\lineskip by #1 \divide\lineskip by #2%
      \lineskiplimit = \normallineskiplimit%
      \multiply\lineskiplimit by #1 \divide\lineskiplimit by #2%
      \parskip = \normalparskip%
      \multiply\parskip by #1 \divide\parskip by #2%
      \abovedisplayskip = \normaldisplayskip%
      \multiply\abovedisplayskip by #1 \divide\abovedisplayskip by #2%
      \belowdisplayskip = \abovedisplayskip%
      \abovedisplayshortskip = \normaldispshortskip%
      \multiply\abovedisplayshortskip by #1%
        \divide\abovedisplayshortskip by #2%
      \belowdisplayshortskip = \abovedisplayshortskip%
      \advance\belowdisplayshortskip by \belowdisplayskip%
      \divide\belowdisplayshortskip by 2%
      \smallskipamount = \skipregister%
      \multiply\smallskipamount by #1 \divide\smallskipamount by #2%
      \medskipamount = \smallskipamount \multiply\medskipamount by 2%
      \bigskipamount = \smallskipamount \multiply\bigskipamount by 4 }
\def\normalbaselines{ \baselineskip=\normalbaselineskip%
   \lineskip=\normallineskip \lineskiplimit=\normallineskip%
   \iftwelv@\else \multiply\baselineskip by 4 \divide\baselineskip by 5%
     \multiply\lineskiplimit by 4 \divide\lineskiplimit by 5%
     \multiply\lineskip by 4 \divide\lineskip by 5 \fi }
\Twelvepoint  
\interlinepenalty=50
\interfootnotelinepenalty=5000
\predisplaypenalty=9000
\postdisplaypenalty=500
\hfuzz=1pt
\vfuzz=0.2pt
%
%
%
\def\pagecontents{%
   \ifvoid\topins\else\unvbox\topins\vskip\skip\topins\fi
   \dimen@ = \dp255 \unvbox255
   \ifvoid\footins\else\vskip\skip\footins\footrule\unvbox\footins\fi
   \ifr@ggedbottom \kern-\dimen@ \vfil \fi }
\def\makeheadline{\vbox to 0pt{ \skip@=\topskip
      \advance\skip@ by -12pt \advance\skip@ by -2\normalbaselineskip
      \vskip\skip@ \line{\vbox to 12pt{}\the\headline} \vss
      }\nointerlineskip}
\def\makefootline{\baselineskip = 1.5\normalbaselineskip
                 \line{\the\footline}}
\newif\iffrontpage
\newif\ifletterstyle
\newif\ifp@genum
\def\nopagenumbers{\p@genumfalse}
\def\pagenumbers{\p@genumtrue}
\pagenumbers
\newtoks\paperheadline
\newtoks\letterheadline
\newtoks\letterfrontheadline
\newtoks\lettermainheadline
\newtoks\paperfootline
\newtoks\letterfootline
\newtoks\date
\footline={\ifletterstyle\the\letterfootline\else\the\paperfootline\fi}
\paperfootline={\hss\iffrontpage\else\ifp@genum\tenrm\folio\hss\fi\fi}
\letterfootline={\hfil}
\headline={\ifletterstyle\the\letterheadline\else\the\paperheadline\fi}
\paperheadline={\hfil}
\letterheadline{\iffrontpage\the\letterfrontheadline
     \else\the\lettermainheadline\fi}
\lettermainheadline={\rm\ifp@genum page \ \folio\fi\hfil\the\date}
\def\monthname{\relax\ifcase\month 0/\or January\or February\or
   March\or April\or May\or June\or July\or August\or September\or
   October\or November\or December\else\number\month/\fi}
\date={\monthname\ \number\day, \number\year}
\countdef\pagenumber=1  \pagenumber=1
\def\advancepageno{\global\advance\pageno by 1
   \ifnum\pagenumber<0 \global\advance\pagenumber by -1
    \else\global\advance\pagenumber by 1 \fi \global\frontpagefalse }
\def\folio{\ifnum\pagenumber<0 \romannumeral-\pagenumber
           \else \number\pagenumber \fi }
\def\footrule{\dimen@=\prevdepth\nointerlineskip
   \vbox to 0pt{\vskip -0.25\baselineskip \hrule width 0.35\hsize \vss}
   \prevdepth=\dimen@ }
\newtoks\foottokens
\foottokens={\Tenpoint\singlespace}
\newdimen\footindent
\footindent=24pt
\def\vfootnote#1{\insert\footins\bgroup  \the\foottokens
   \interlinepenalty=\interfootnotelinepenalty \floatingpenalty=20000
   \splittopskip=\ht\strutbox \boxmaxdepth=\dp\strutbox
   \leftskip=\footindent \rightskip=\z@skip
   \parindent=0.5\footindent \parfillskip=0pt plus 1fil
   \spaceskip=\z@skip \xspaceskip=\z@skip
   \Textindent{$ #1 $}\footstrut\futurelet\next\fo@t}
\def\Textindent#1{\noindent\llap{#1\enspace}\ignorespaces}
\def\footnote#1{\attach{#1}\vfootnote{#1}}

\let\footsymbol=\star
\newcount\lastf@@t           \lastf@@t=-1
\newcount\footsymbolcount    \footsymbolcount=0
\newif\ifPhysRev
\def\footsymbolgen{\relax \ifPhysRev \iffrontpage \NPsymbolgen\else
      \PRsymbolgen\fi \else \NPsymbolgen\fi
   \global\lastf@@t=\pageno \footsymbol }
\def\NPsymbolgen{\ifnum\footsymbolcount<0 \global\footsymbolcount=0\fi
   {\iffrontpage \else \advance\lastf@@t by 1 \fi
    \ifnum\lastf@@t<\pageno \global\footsymbolcount=0
     \else \global\advance\footsymbolcount by 1 \fi }
   \ifcase\footsymbolcount \fd@f\star\or \fd@f\dagger\or \fd@f\ddagger\or
    \fd@f\ast\or \fd@f\natural\or \fd@f\diamond\or \fd@f\bullet\or
    \fd@f\nabla\else \fd@f\dagger\global\footsymbolcount=0 \fi }
\def\fd@f#1{\xdef\footsymbol{#1}}
\def\PRsymbolgen{\ifnum\footsymbolcount>0 \global\footsymbolcount=0\fi
      \global\advance\footsymbolcount by -1
      \xdef\footsymbol{\sharp\number-\footsymbolcount} }
\def\space@ver#1{\let\@sf=\empty \ifmmode #1\else \ifhmode
   \edef\@sf{\spacefactor=\the\spacefactor}\unskip${}#1$\relax\fi\fi}
\def\attach#1{\space@ver{\strut^{\mkern 2mu #1} }\@sf\ }
%
%
%
\newcount\chapternumber      \chapternumber=0
\newcount\sectionnumber      \sectionnumber=0
\newcount\equanumber         \equanumber=0
\let\chapterlabel=0
\newtoks\chapterstyle        \chapterstyle={\Number}
\newskip\chapterskip         \chapterskip=\bigskipamount
\newskip\sectionskip         \sectionskip=\medskipamount
\newskip\headskip            \headskip=8pt plus 3pt minus 3pt
\newdimen\chapterminspace    \chapterminspace=15pc
\newdimen\sectionminspace    \sectionminspace=10pc
\newdimen\referenceminspace  \referenceminspace=25pc
\def\chapterreset{\global\advance\chapternumber by 1
   \ifnum\equanumber<0 \else\global\equanumber=0\fi
   \sectionnumber=0 \makel@bel}
\def\makel@bel{\xdef\chapterlabel{%
\the\chapterstyle{\the\chapternumber}.}}
\def\sectionlabel{\number\sectionnumber \quad }
\def\alphabetic#1{\count255='140 \advance\count255 by #1\char\count255}
\def\Alphabetic#1{\count255='100 \advance\count255 by #1\char\count255}
\def\Roman#1{\uppercase\expandafter{\romannumeral #1}}
\def\roman#1{\romannumeral #1}
\def\Number#1{\number #1}
\def\unnumberedchapters{\let\makel@bel=\relax \let\chapterlabel=\relax
\let\sectionlabel=\relax \equanumber=-1 }
\def\titlestyle#1{\par\begingroup \interlinepenalty=9999
     \leftskip=0.02\hsize plus 0.23\hsize minus 0.02\hsize
     \rightskip=\leftskip \parfillskip=0pt
     \hyphenpenalty=9000 \exhyphenpenalty=9000
     \tolerance=9999 \pretolerance=9000
     \spaceskip=0.333em \xspaceskip=0.5em
     \iftwelv@\twelvepoint\fourteenrm\else\twelvepoint\fi
   \noindent #1\par\endgroup }
\def\spacecheck#1{\dimen@=\pagegoal\advance\dimen@ by -\pagetotal
   \ifdim\dimen@<#1 \ifdim\dimen@>0pt \vfil\break \fi\fi}
\def\chapter#1{\par \penalty-300 \vskip\chapterskip
   \spacecheck\chapterminspace
   \chapterreset \titlestyle{\chapterlabel \ #1}
   \nobreak\vskip\headskip \penalty 30000
   \wlog{\string\chapter\ \chapterlabel} }

\def\section#1{\par \ifnum\the\lastpenalty=30000\else
   \penalty-200\vskip\sectionskip \spacecheck\sectionminspace\fi
   \wlog{\string\section\ \chapterlabel \the\sectionnumber}
   \global\advance\sectionnumber by 1  \noindent
   {\caps\enspace\chapterlabel \sectionlabel #1}\par
   \nobreak\vskip\headskip \penalty 30000 }
\def\subsection#1{\par
   \ifnum\the\lastpenalty=30000\else \penalty-100\smallskip \fi
   \noindent\undertext{#1}\enspace \vadjust{\penalty5000}}

\def\undertext#1{\vtop{\hbox{#1}\kern 1pt \hrule}}
\def\APPENDIX#1#2{\par\penalty-300\vskip\chapterskip
   \spacecheck\chapterminspace \chapterreset \xdef\chapterlabel{#1}
   \titlestyle{APPENDIX #2} \nobreak\vskip\headskip \penalty 30000
   \wlog{\string\Appendix\ \chapterlabel} }
\def\Appendix#1{\APPENDIX{#1}{#1}}
\def\appendix{\APPENDIX{A}{}}
%
%
%

\newif\ifdraftmode
\draftmodefalse

\def\eqname#1{\relax \ifnum\equanumber<0
     \xdef#1{{(\number-\equanumber)}}\global\advance\equanumber by -1
    \else \global\advance\equanumber by 1
      \xdef#1{{(\chapterlabel \number\equanumber)}} \fi}
\def\eqn#1{\eqno\eqname{#1}#1\ifdraftmode\rlap{\sevenrm\ \string#1}\fi}


%
\def\eqinsert#1{\noalign{\dimen@=\prevdepth \nointerlineskip
   \setbox0=\hbox to\displaywidth{\hfil #1}
   \vbox to 0pt{\vss\hbox{$\!\box0\!$}\kern-0.5\baselineskip}
   \prevdepth=\dimen@}}
%

%

%

%
%
\def\GENITEM#1;#2{\par \hangafter=0 \hangindent=#1
    \Textindent{$ #2 $}\ignorespaces}
\outer\def\newitem#1=#2;{\gdef#1{\GENITEM #2;}}
\newdimen\itemsize                \itemsize=30pt
\newitem\item=1\itemsize;
\newitem\sitem=1.75\itemsize;     
\newitem\ssitem=2.5\itemsize;     
\outer\def\newlist#1=#2&#3&#4;{\toks0={#2}\toks1={#3}%
   \count255=\escapechar \escapechar=-1
   \alloc@0\list\countdef\insc@unt\listcount     \listcount=0
   \edef#1{\par
      \countdef\listcount=\the\allocationnumber
      \advance\listcount by 1
      \hangafter=0 \hangindent=#4
      \Textindent{\the\toks0{\listcount}\the\toks1}}
   \expandafter\expandafter\expandafter
    \edef\c@t#1{begin}{\par
      \countdef\listcount=\the\allocationnumber \listcount=1
      \hangafter=0 \hangindent=#4
      \Textindent{\the\toks0{\listcount}\the\toks1}}
   \expandafter\expandafter\expandafter
    \edef\c@t#1{con}{\par \hangafter=0 \hangindent=#4 \noindent}
   \escapechar=\count255}
\def\c@t#1#2{\csname\string#1#2\endcsname}
\newlist\point=\Number&.&1.0\itemsize;
\newlist\subpoint=(\alphabetic&)&1.75\itemsize;
\newlist\subsubpoint=(\roman&)&2.5\itemsize;
\let\spoint=\subpoint

%
%
%
\newcount\referencecount     \referencecount=0
\newif\ifreferenceopen       \newwrite\referencewrite
\newtoks\rw@toks
\def\NPrefmark#1{\attach{\scriptscriptstyle [ #1 ] }}
\let\PRrefmark=\attach
\def\refmark#1{\relax\ifPhysRev\PRrefmark{#1}\else\NPrefmark{#1}\fi}
\def\refend{\refmark{\number\referencecount}}
\newcount\lastrefsbegincount \lastrefsbegincount=0
\def\refsend{\refmark{\count255=\referencecount
   \advance\count255 by-\lastrefsbegincount
   \ifcase\count255 \number\referencecount
   \or \number\lastrefsbegincount,\number\referencecount
   \else \number\lastrefsbegincount-\number\referencecount \fi}}
\def\refch@ck{\chardef\rw@write=\referencewrite
   \ifreferenceopen \else \referenceopentrue
   \immediate\openout\referencewrite=referenc.tex \fi}
%
{\catcode`\^^M=\active 
  \gdef\obeyendofline{\catcode`\^^M\active \let^^M\ }}%
%
{\catcode`\^^M=\active 
  \gdef\ignoreendofline{\catcode`\^^M=5}}
{\obeyendofline\gdef\rw@start#1{\def\t@st{#1} \ifx\t@st\blankend%
\endgroup \@sf \relax \else \ifx\t@st\bl@nkend \endgroup \@sf \relax%
\else \rw@begin#1
\backtotext
\fi \fi } }
{\obeyendofline\gdef\rw@begin#1
{\def\n@xt{#1}\rw@toks={#1}\relax%
\rw@next}}
\def\blankend{}
{\obeylines\gdef\bl@nkend{
}}
\newif\iffirstrefline  \firstreflinetrue
\def\rwr@teswitch{\ifx\n@xt\blankend \let\n@xt=\rw@begin %
 \else\iffirstrefline \global\firstreflinefalse%
\immediate\write\rw@write{\noexpand\obeyendofline \the\rw@toks}%
\let\n@xt=\rw@begin%
      \else\ifx\n@xt\rw@@d \def\n@xt{\immediate\write\rw@write{%
        \noexpand\ignoreendofline}\endgroup \@sf}%
             \else \immediate\write\rw@write{\the\rw@toks}%
             \let\n@xt=\rw@begin\fi\fi \fi}
\def\rw@next{\rwr@teswitch\n@xt}
\def\rw@@d{\backtotext} \let\rw@end=\relax
\let\backtotext=\relax

\newdimen\refindent     \refindent=30pt
\def\refitem#1{\par \hangafter=0 \hangindent=\refindent \Textindent{#1}}
\def\REFNUM#1{\space@ver{}\refch@ck \firstreflinetrue%
 \global\advance\referencecount by 1 \xdef#1{\the\referencecount}}
\def\refnum#1{\space@ver{}\refch@ck \firstreflinetrue%
 \global\advance\referencecount by 1 \xdef#1{\the\referencecount}\refend}

\def\REF#1{\REFNUM#1%
 \immediate\write\referencewrite{%
            \noexpand\refitem{\ifdraftmode{\sevenrm 
               \noexpand\string\string#1\ }\fi#1.}}%
      \begingroup\obeyendofline\rw@start}
\def\ref{\refnum\?%
 \immediate\write\referencewrite{\noexpand\refitem{\?.}}%
\begingroup\obeyendofline\rw@start}
\def\Ref#1{\refnum#1%
 \immediate\write\referencewrite{\noexpand\refitem{#1.}}%
\begingroup\obeyendofline\rw@start}
\def\REFS#1{\REFNUM#1\global\lastrefsbegincount=\referencecount
\immediate\write\referencewrite{\noexpand\refitem{#1.}}%
\begingroup\obeyendofline\rw@start}
%

%
%

%
\newcount\figurecount     \figurecount=0
\newif\iffigureopen       \newwrite\figurewrite
\def\figch@ck{\chardef\rw@write=\figurewrite \iffigureopen\else
   \immediate\openout\figurewrite=figures.aux
   \figureopentrue\fi}
\def\FIGNUM#1{\space@ver{}\figch@ck \firstreflinetrue%
 \global\advance\figurecount by 1 \xdef#1{\the\figurecount}}
\def\FIG#1{\FIGNUM#1
   \immediate\write\figurewrite{\noexpand\refitem{#1.}}%
   \begingroup\obeyendofline\rw@start}
\def\fig{\FIGNUM\? fig.~\?%
\immediate\write\figurewrite{\noexpand\refitem{\?.}}%
\begingroup\obeyendofline\rw@start}
\def\figure{\FIGNUM\? figure~\?
   \immediate\write\figurewrite{\noexpand\refitem{\?.}}%
   \begingroup\obeyendofline\rw@start}
\def\Fig{\FIGNUM\? Fig.~\?%
\immediate\write\figurewrite{\noexpand\refitem{\?.}}%
\begingroup\obeyendofline\rw@start}
\def\Figure{\FIGNUM\? Figure~\?%
\immediate\write\figurewrite{\noexpand\refitem{\?.}}%
\begingroup\obeyendofline\rw@start}
\newcount\tablecount     \tablecount=0
\newif\iftableopen       \newwrite\tablewrite
\def\tabch@ck{\chardef\rw@write=\tablewrite \iftableopen\else
   \immediate\openout\tablewrite=tables.aux
   \tableopentrue\fi}
\def\TABNUM#1{\space@ver{}\tabch@ck \firstreflinetrue%
 \global\advance\tablecount by 1 \xdef#1{\the\tablecount}}
\def\TABLE#1{\TABNUM#1
   \immediate\write\tablewrite{\noexpand\refitem{#1.}}%
   \begingroup\obeyendofline\rw@start}
\def\Table{\TABNUM\? Table~\?%
\immediate\write\tablewrite{\noexpand\refitem{\?.}}%
\begingroup\obeyendofline\rw@start}
\newif\ifsymbolopen       \newwrite\symbolwrite
\def\symch@ck{\ifsymbolopen\else
   \immediate\openout\symbolwrite=symbols.aux
   \symbolopentrue\fi}
\def\symdef#1#2{\def#1{#2}%
      \symch@ck%
      \immediate\write\symbolwrite{$$ \hbox{\noexpand\string\string#1} 
                \noexpand\qquad\noexpand\longrightarrow\noexpand\qquad 
                           \string#1 $$}}
%
%

%
%
%
\def\masterreset{\global\pagenumber=1 \global\chapternumber=0
   \global\equanumber=0 \global\sectionnumber=0
   \global\referencecount=0 \global\figurecount=0 \global\tablecount=0 }
\def\FRONTPAGE{\ifvoid255\else\vfill\penalty-2000\fi
      \masterreset\global\frontpagetrue
      \global\lastf@@t=0 \global\footsymbolcount=0}

\def\paperstyle{\letterstylefalse\normalspace\papersize}
\def\letterstyle{\letterstyletrue\singlespace\lettersize}
\def\papersize{\hsize=35pc\vsize=50pc\hoffset=1pc\voffset=6pc
               \skip\footins=\bigskipamount}
\def\lettersize{\hsize=6.5in\vsize=8.5in\hoffset=0in\voffset=1in
   \skip\footins=\smallskipamount \multiply\skip\footins by 3 }
\paperstyle   
%
%
\def\MEMO{\letterstyle\FRONTPAGE \letterfrontheadline={\hfil}
    \line{\quad\fourteenrm NTC MEMORANDUM\hfil\twelverm\the\date\quad}
    \medskip \memod@f}

\def\memit@m#1{\smallskip \hangafter=0 \hangindent=1in
      \Textindent{\caps #1}}
\def\memod@f{\xdef\to{\memit@m{To:}}\xdef\from{\memit@m{From:}}%
     \xdef\topic{\memit@m{Topic:}}\xdef\subject{\memit@m{Subject:}}%
     \xdef\rule{\bigskip\hrule height 1pt\bigskip}}
\memod@f
\newskip\lettertopfil
\lettertopfil = 0pt plus 1.5in minus 0pt
\newskip\letterbottomfil
\letterbottomfil = 0pt plus 2.3in minus 0pt
\newskip\spskip \setbox0\hbox{\ } \spskip=-1\wd0
\def\addressee#1{\medskip\rightline{\the\date\hskip 30pt} \bigskip
   \vskip\lettertopfil
   \ialign to\hsize{\strut ##\hfil\tabskip 0pt plus \hsize \cr #1\crcr}
   \medskip\noindent\hskip\spskip}
\newskip\signatureskip       \signatureskip=40pt
\def\signed#1{\par \penalty 9000 \bigskip \dt@pfalse
  \everycr={\noalign{\ifdt@p\vskip\signatureskip\global\dt@pfalse\fi}}
  \setbox0=\vbox{\singlespace \halign{\tabskip 0pt \strut ##\hfil\cr
   \noalign{\global\dt@ptrue}#1\crcr}}
  \line{\hskip 0.5\hsize minus 0.5\hsize \box0\hfil} \medskip }

\def\endletter{\ifnum\pagenumber=1 \vskip\letterbottomfil\supereject
\else \vfil\supereject \fi}
\newbox\letterb@x
\def\lettertext{\par\unvcopy\letterb@x\par}
\def\multiletter{\setbox\letterb@x=\vbox\bgroup
      \everypar{\vrule height 1\baselineskip depth 0pt width 0pt }
      \singlespace \topskip=\baselineskip }
\def\letterend{\par\egroup}
%
%
%
\newskip\frontpageskip
\newtoks\pubtype
\newtoks\Pubnum
\newtoks\pubnum
\newif\ifp@bblock  \p@bblocktrue
\def\PH@SR@V{\doubl@true \baselineskip=24.1pt plus 0.2pt minus 0.1pt
             \parskip= 3pt plus 2pt minus 1pt }
\def\PHYSREV{\paperstyle\PhysRevtrue\PH@SR@V}
\def\titlepage{\FRONTPAGE\paperstyle\ifPhysRev\PH@SR@V\fi
   \ifp@bblock\p@bblock\fi}
\def\nopubblock{\p@bblockfalse}
\def\endpage{\vfil\break}
\frontpageskip=1\medskipamount plus .5fil
\pubtype={\tensl Preliminary Version}
\Pubnum={$\caps SLAC - PUB - \the\pubnum $}
\pubnum={0000}
\def\p@bblock{\begingroup \tabskip=\hsize minus \hsize
   \baselineskip=1.5\ht\strutbox \topspace-2\baselineskip
   \halign to\hsize{\strut ##\hfil\tabskip=0pt\crcr
   \the\Pubnum\cr \the\date\cr \the\pubtype\cr}\endgroup}
\def\title#1{\vskip\frontpageskip \titlestyle{#1} \vskip\headskip }
\def\author#1{\vskip\frontpageskip\titlestyle{\twelvecp #1}\nobreak}

\def\address#1{\par\kern 5pt\titlestyle{\twelvepoint\it #1}}
\def\andaddress{\par\kern 5pt \centerline{\sl and} \address}

\def\abstract{\vskip\frontpageskip\centerline{\fourteenrm ABSTRACT}
              \vskip\headskip }

%
%
%

\def\\{\relax\ifmmode\backslash\else$\backslash$\fi}
\def\globaleqnumbers{\relax\if\equanumber<0\else\global\equanumber=-1\fi}

\def\journal#1&#2(#3){\unskip, \sl #1~\bf #2 \rm (19#3) }

\def\topspace{\hrule height 0pt depth 0pt \vskip}

\let\int=\intop         
\def\prop{\mathrel{{\mathchoice{\pr@p\scriptstyle}{\pr@p\scriptstyle}{
                \pr@p\scriptscriptstyle}{\pr@p\scriptscriptstyle} }}}
\def\pr@p#1{\setbox0=\hbox{$\cal #1 \char'103$}
   \hbox{$\cal #1 \char'117$\kern-.4\wd0\box0}}
\def\lsim{\mathrel{\mathpalette\@versim<}}
\def\gsim{\mathrel{\mathpalette\@versim>}}
\def\@versim#1#2{\lower0.5ex\vbox{\baselineskip\z@skip\lineskip-.1ex
  \lineskiplimit\z@\ialign{$\m@th#1\hfil##\hfil$\crcr#2\crcr\sim\crcr}}}
%
%
%
\let\sec@nt=\sec
\def\sec{\relax\ifmmode\let\n@xt=\sec@nt\else\let\n@xt\section\fi\n@xt}
\def\obsolete#1{\message{Macro \string #1 is obsolete.}}
\def\firstsec#1{\obsolete\firstsec \section{#1}}
\def\firstsubsec#1{\obsolete\firstsubsec \subsection{#1}}
\def\thispage#1{\obsolete\thispage \global\pagenumber=#1\frontpagefalse}
\def\thischapter#1{\obsolete\thischapter \global\chapternumber=#1}
\def\nextequation#1{\obsolete\nextequation \global\equanumber=#1
   \ifnum\the\equanumber>0 \global\advance\equanumber by 1 \fi}
\def\BOXITEM{\afterassigment\B@XITEM\setbox0=}
\def\B@XITEM{\par\hangindent\wd0 \noindent\box0 }
%

%
%
%
%
%
\lock
\message{   }
%
%
%












\newbox\figbox
\newdimen\zero  \zero=0pt
\newdimen\figmove
\newdimen\figwidth
\newdimen\figheight
\newdimen\textwidth
\newtoks\figtoks
\newcount\figcounta
\newcount\figcountb
\newcount\figlines
\def\figreset{\global\figcounta=-1 \global\figcountb=-1
\global\figmove=\baselineskip
\global\figlines=1 \global\figtoks={ } }
\def\picture#1by#2:#3{\global\setbox\figbox=\vbox{\vskip #1
\hbox{\vbox{\hsize=#2 \noindent #3}}}
\global\setbox\figbox=\vbox{\kern 10pt
\hbox{\kern 10pt \box\figbox \kern 10pt }\kern 10pt}
\global\figwidth=1\wd\figbox
\global\figheight=1\ht\figbox
\global\textwidth=\hsize
\global\advance\textwidth by - \figwidth }
\def\figtoksappend{\edef\temp##1{\global\figtoks=%
{\the\figtoks ##1}}\temp}
\def\figparmsa#1{\loop \global\advance\figcounta by 1
\ifnum \figcounta < #1
\figtoksappend{ 0pt \the\hsize }
\global\advance\figlines by 1
\repeat }
\def\figparmsb#1{\loop \global\advance\figcountb by 1
\ifnum \figcountb < #1
\figtoksappend{ \the\figwidth \the\textwidth}
\global\advance\figlines by 1
\repeat }
\def\figtext#1:#2:#3{\figreset%
\figparmsa{#1}%
\figparmsb{#2}%
\multiply\figmove by #1%
\global\setbox\figbox=\vbox to 0pt{\vskip \figmove  \hbox{\box\figbox}
\vss }
\parshape=\the\figlines\the\figtoks\the\zero\the\hsize
\noindent
\rlap{\box\figbox} #3}
\def\Buildrel#1\under#2{\mathrel{\mathop{#2}\limits_{#1}}}
\def\llongrarrow{\hbox to 40pt{\rightarrowfill}}



\def\boxit#1{\vbox{\hrule\hbox{\vrule\kern3pt
\vbox{\kern3pt#1\kern3pt}\kern3pt\vrule}\hrule}}
\newdimen\str
\def\fboxit#1#2{\vbox{\hrule height #1 \hbox{\vrule width #1
\kern3pt \vbox{\kern3pt#2\kern3pt}\kern3pt \vrule width #1 }
\hrule height #1 }}
\def\tran#1#2{\transpoint \hfuzz 5pt \gdef\label{#1}
\vbox to \the\vsize{\hsize \the\hsize #2} \par \eject }
\newdimen\baseskip
\newdimen\lskip
\lskip=\lineskip
\newdimen\transize
\newdimen\tall
\def\transpoint{\gdef\rm{\fam0\eighteenrm}
\font\twentyfourit = cmti10 scaled \magstep5
\font\twentyfourrm = cmr10 scaled \magstep5
\font\twentyfourbf = cmbx10 scaled \magstep5
\font\twentyeightsy = cmsy10 scaled \magstep5
\font\eighteenrm = cmr10 scaled \magstep3
\font\eighteenb = cmbx10 scaled \magstep3
\font\eighteeni = cmmi10 scaled \magstep3
\font\eighteenit = cmti10 scaled \magstep3
\font\eighteensl = cmsl10 scaled \magstep3
\font\eighteensy = cmsy10 scaled \magstep3
\font\eighteencaps = cmr10 scaled \magstep3
\font\eighteenmathex = cmex10 scaled \magstep3
\font\fourteenrm=cmr10 scaled \magstep2
\font\fourteeni=cmmi10 scaled \magstep2
\font\fourteenit = cmti10 scaled \magstep2
\font\fourteensy=cmsy10 scaled \magstep2
\font\fourteenmathex = cmex10 scaled \magstep2
\parindent 20pt
\global\hsize = 7.0in
\global\vsize = 8.9in
\dimen\transize = \the\hsize
\dimen\tall = \the\vsize
\def\sy{\eighteensy }
\def\sl{\eighteens }
\def\bf{\eighteenb }
\def\it{\eighteenit }
\def\caps{\eighteencaps }
\textfont0=\eighteenrm \scriptfont0=\fourteenrm
\scriptscriptfont0=\twelverm
\textfont1=\eighteeni \scriptfont1=\fourteeni \scriptscriptfont1=\twelvei
\textfont2=\eighteensy \scriptfont2=\fourteensy
\scriptscriptfont2=\twelvesy
\textfont3=\eighteenmathex \scriptfont3=\eighteenmathex
\scriptscriptfont3=\eighteenmathex
\global\baselineskip 35pt
\global\lineskip 15pt
\global\parskip 5pt  plus 1pt minus 1pt
\global\abovedisplayskip  3pt plus 10pt minus 10pt
\global\belowdisplayskip 3pt plus 10pt minus 10pt
\def\rtitle##1{\centerline{\undertext{\twentyfourrm ##1}}}
\def\ititle##1{\centerline{\undertext{\twentyfourit ##1}}}
\def\ctitle##1{\centerline{\undertext{\caps ##1}}}
\def\vstrut{\hbox{\vrule width 0pt height .35in depth .15in }}
\def\cline##1{\centerline{\vstrut ##1}}
\output{\shipout\vbox{\vskip .5in
\pagecontents \vfill
\hbox to \the\hsize{\hfill{\tenbf \label} } }
\global\advance\count0 by 1 }
\rm }


%
%
%

%

%

%

%
{\obeyspaces\global\let =\ }
%
%
%
\widowpenalty 1000
\thickmuskip 4mu plus 4mu

\def\bspac{\noalign{\vskip 4pt}}

\unlock
%
\Pubnum={${\twelverm IU/NTC}\  \the\pubnum $}
\pubnum={0000}
\def\p@nnlock{\begingroup \tabskip=\hsize minus \hsize
   \baselineskip=1.5\ht\strutbox \topspace-2\baselineskip
   \noindent\strut\the\Pubnum \hfill \the\date   \endgroup}
\def\titlepage{\FRONTPAGE\paperstyle\p@nnlock}
\def\displaylines#1{\displ@y
  \halign{\hbox to\displaywidth{$\hfil\displaystyle##\hfil$}\crcr
    #1\crcr}}
\def\addressee#1{\null
   \bigskip\medskip\rightline{\the\date\hskip 30pt}
   \vskip\lettertopfil
   \ialign to\hsize{\strut ##\hfil\tabskip 0pt plus \hsize \cr #1\crcr}
   \medskip\vskip 3pt\noindent}
\def\tmsaddressee#1#2{
   \vskip\lettertopfil
  \setbox0=\vbox{\singlespace \halign{\tabskip 0pt \strut ##\hfil\cr
   \noalign{\global\dt@ptrue}#1\crcr}}
  \line{\hskip 0.7\hsize minus 0.7\hsize \box0\hfil}
   \bigskip
   \vskip .2in
   \ialign to\hsize{\strut ##\hfil\tabskip 0pt plus \hsize \cr #2\crcr}
   \medskip\vskip 3pt\noindent}
\def\makeheadline{\vbox to 0pt{ \skip@=\topskip
      \advance\skip@ by -12pt \advance\skip@ by -2\normalbaselineskip
      \vskip\skip@  \vss
      }\nointerlineskip}
\def\signed#1{\par \penalty 9000 \bigskip \vskip .06in\dt@pfalse
  \everycr={\noalign{\ifdt@p\vskip\signatureskip\global\dt@pfalse\fi}}
  \setbox0=\vbox{\singlespace \halign{\tabskip 0pt \strut ##\hfil\cr
   \noalign{\global\dt@ptrue}#1\crcr}}
  \line{\hskip 0.5\hsize minus 0.5\hsize \box0\hfil} \medskip }
\def\lettersize{\hsize=6.25in\vsize=8.5in\hoffset=0in\voffset=1in
   \skip\footins=\smallskipamount \multiply\skip\footins by 3 }
%
%
%
%
%
\outer\def\newnewlist#1=#2&#3&#4&#5;{\toks0={#2}\toks1={#3}%
   \dimen1=\hsize  \advance\dimen1 by -#4
   \dimen2=\hsize  \advance\dimen2 by -#5
   \count255=\escapechar \escapechar=-1
   \alloc@0\list\countdef\insc@unt\listcount     \listcount=0
   \edef#1{\par
      \countdef\listcount=\the\allocationnumber
      \advance\listcount by 1
      \parshape=2 #4 \dimen1 #5 \dimen2
      \Textindent{\the\toks0{\listcount}\the\toks1}}
   \expandafter\expandafter\expandafter
    \edef\c@t#1{begin}{\par
      \countdef\listcount=\the\allocationnumber \listcount=1
      \parshape=2 #4 \dimen1 #5 \dimen2
      \Textindent{\the\toks0{\listcount}\the\toks1}}
   \expandafter\expandafter\expandafter
    \edef\c@t#1{con}{\par \parshape=2 #4 \dimen1 #5 \dimen2 \noindent}
   \escapechar=\count255}
\def\c@t#1#2{\csname\string#1#2\endcsname}
%
%
%
%
%
%
%
\def\noparGENITEM#1;{\hangafter=0 \hangindent=#1
    \ignorespaces\noindent}
\outer\def\noparnewitem#1=#2;{\gdef#1{\noparGENITEM #2;}}
\noparnewitem\spoint=1.5\itemsize;
%
%
%
\def\MEMO{\letterstyle\FRONTPAGE \letterfrontheadline={\hfil}
      \hoffset=1in \voffset=1.21in
    \line{\hskip .8in  \special{overlay ntcmemo.dat}
          \quad\fourteenrm NTC MEMORANDUM\hfil\twelverm\the\date\quad}
    \medskip\medskip \memod@f}

\def\memit@m#1{\smallskip \hangafter=0 \hangindent=1in
      \Textindent{\caps #1}}
\def\memod@f{\xdef\to{\memit@m{To:}}\xdef\from{\memit@m{From:}}%
     \xdef\topic{\memit@m{Topic:}}\xdef\subject{\memit@m{Subject:}}%
     \xdef\rule{\bigskip\hrule height 1pt\bigskip}}
\memod@f
\lock
%

%
%
%
%
%
%
%

\let\ssize=\scriptstyle
\let\sssize\scriptscriptstyle
%
%
%
%
%
\def\tildesymbol{\mathchar"0218 }
\def\undervec#1{\setbox0\hbox{$\tildesymbol$}\setbox1\hbox{$#1$}#1%
                \dimen0=\wd0\advance\dimen0by\wd1\divide\dimen0by2
                 \kern-\dimen0\lower1.3ex\hbox{$\tildesymbol$}}
\def\sundervec#1{\setbox0\hbox{$\ssize\tildesymbol$}
                 \setbox1\hbox{$\ssize #1$}#1%
                \dimen0=\wd0\advance\dimen0by\wd1\divide\dimen0by2
                 \kern-\dimen0\lower.85ex\hbox{$\ssize\tildesymbol$}}
\def\ssundervec#1{\setbox0\hbox{$\sssize\tildesymbol$}
                 \setbox1\hbox{$\sssize #1$}#1%
                \dimen0=\wd0\advance\dimen0by\wd1\divide\dimen0by2
                 \kern-\dimen0\lower.6ex\hbox{$\sssize\tildesymbol$}}
%

%


%

%

%

%

%

%

%

%
%
%
%
%

%

%

%

%

%

%
%
%
%
%
%

%

%

%

%

%
%
%
%
%
\def\intback{\kern-.1em}

%
%
%
%
%
\def\threej#1#2#3#4#5#6{\left(\matrix{#1 & #2 & #3 \cr
                                      #4 & #5 & #6 \cr}\right)}
\def\sixj#1#2#3#4#5#6{\left\{\matrix{#1 & #2 & #3 \cr
                                      #4 & #5 & #6 \cr}\right\}}

\def\bspac{\noalign{\vskip 4pt}}

%
%
%
%
%

%
%
%
%
%

%
%
%
%
%

%

%
%

%

%

%

%

%

%

%

%

%
%
%
%
%

%
%
%
%

%
%
%
\def\papersize{\hsize=6.25in\vsize=8.60in\hoffset=0.0in\voffset=0.1in
                \skip\footins=\bigskipamount}
\paperstyle
\doublespace
\pretolerance=500
\tolerance=500
\widowpenalty 5000
\thinmuskip=3mu
\medmuskip=4mu plus 2mu minus 4mu
\thickmuskip=5mu plus 5mu
\def\NPrefmark#1{\attach{\ssize #1{}}}
%
%
%
%
\def\NPrefmark#1{\attach{\scriptstyle  #1 }}

\def\Llo{{\lambda_1}}
\def\Llt{{\lambda_2}}
\def\Llth{{\lambda_3}}
\def\Llf{{\lambda_4}}

\PHYSREV
%
%

%
\REF\TAFFAR{R. Anni and L. Taffara, Nuovo Cimento {\bf 22A}, 12 (1974).}

\REF\ELBAZM{E. Elbaz, J. Meyer and R. Nahabetian, Lett. Nuovo Cimento
{\bf 10}, 418 (1974).}

\REF\BAEZA{A. Baeza, B. Bilwes, R. Bilwes, J. Dias, J.L. Ferrero 
and J. Raynal, Nucl. Phys. A {\bf 437}, 93 (1985).} 

\REF\WHITE{G. D. White, K. T. R. Davies and P. J. Siemens, Ann. Phys. 
   (NY) {\bf 187}, 198 (1988).}

\REF\DAVIES{K. T. R. Davies, M. R. Strayer and G. D. White, J. Phys. 
   G {\bf 14}, 961 (1988).}

\REF\DAVIEST{K. T. R. Davies, J. Phys. 
  G {\bf 14}, 973 (1988).}

\REF\MEHR{R. Mehrem, J.T. Londergan and M.~H.~Macfarlane, J. Phys. A {\bf 24}, 1435 (1991).}

\REF\MEHREM{ R. Mehrem, {\it Ph.D. Thesis}, (Indiana University 1992)(unpublished).}

\REF\MEHRE{R. Mehrem, J.T. Londergan and G.~E.~Walker, Phys. Rev. C {\bf 48}, 1192 (1993).}

\REF\MEH{R. Mehrem, Appl. Math. Comp. {\bf 217}, 5360 (2011), 
arXiv: math-ph/0909.0494, 2010.}

\REF\CHEN{J. Chen and J. Su, Phys. Rev. D {\bf 69}, 076003 (2004).}

\REF\HOH{A. Hohenegger, Phys. Rev. D {\bf 79}, 063502 (2009), arXiv:0806.3098.}

\REF\HOHDIS{A.~Hohenegger, {\it Ph.D.~Thesis}, (University of Heidelberg, 2010).}

\REF\JORISA{J.~Van Deun and R.~Cools, ACM Trans.~Math.~Softw.~{\bf 32}, 580 (2006).}

\REF\COS{C. Crucean, arXiv: math-ph/0912.3659, 2009.}

\REF\MEHHOH{R.~Mehrem and A.~Hohenegger, J. Phys. A {\bf 43}, 455204 (2010), arXiv: math-ph/1006.2108, 2010.}

\REF\BRNKSA{\rm D.M. Brink and G.R. Satchler, {\it Angular Momentum}
(Oxford University Press, London 1962).}

\REF\EDMNDS{\rm A. R. Edmonds, {\it Angular Momentum in Quantum Mechanics}
(Princeton University Press, Princeton, 1957).}

\REF\GRADST{I.S. Gradshteyn and I.M. Rhyzik: {\it Table of Integrals, 
Series and Products} (Academic Press, New York, 1994).} 

\title{\bf ANALYTICAL EVALUATION OF AN INFINITE INTEGRAL OVER FOUR SPHERICAL BESSEL FUNCTIONS}
\author{R.\ Mehrem$^{\bf a}$} 
\address{\it Associate Lecturer \break
The Open University in the North West \break
351 Altrincham Road \break
Sharston, Manchester M22 4UN \break
United Kingdom}

\footnote{}{\bf a Email: rami.mehrem@btopenworld.com.} 
\endpage
\abstract{An infinite integral over four spherical Bessel functions
of the form 

$\int_0^\infty\,r^2\,j_\Llo(k_1r)\,j_\Llt(k_2r)\,
j_\Llth(k_1r)\,j_\Llf(k_2r)\,dr$ is analytically evaluated. The result involves a finite sum over an associated Legendre function
of integer degree and half integer order with a real argument greater
than 1. Evaluation of such functions is discussed. }
 
\endpage
%
%
\chapter{Introduction}
Infinite integrals over spherical Bessel functions have always been
of interest due to their occurence in nuclear physics 
[\TAFFAR-\MEH], particle physics [\CHEN] and astrophysics 
[\HOH-\HOHDIS], to mention a few. In this paper, an infinite integral over four spherical Bessel functions of the form
$$\int_0^\infty \,r^{2}\,j_\Llo(k_1r)\,j_\Llt(k_2r)\,
j_\Llth(k_3r)\,j_\Llf(k_4r)\,dr\eqn\ONEPONE$$
is analytically evaluated for the special case when $k_3\,=\,k_1$ and $k_4\,=\,k_2$. The parameters $\Llo$, 
$\Llt$, $\Llth$ and $\Llf$ are assumed positive integers.
Recently, a generalised form of the integral has been 
attempted numerically [\JORISA], and analytically [\COS]. However, 
the analytical evaluation involves
a complicated hypergeometric function. In our evaluation, we start by assuming the momenta $k_1$, $k_2$, $k_3$ and $k_4$ to be positive
and form the sides of a quadrilateral, which is a consequence of the conservation of linear momentum in any physical application.
The general result of evaluating eq. $\ONEPONE$ is in 
the form of a finite sum over 3j and 6j symbols with a finite 
integral over Legendre functions with a rational function. This integral is evaluated for the special case when $k_3$=$k_1$ and $k_4$=$k_2$ in appendix A. The final result for this special case involves a finite sum over an associated Legendre function of integer degree, half-integer order and a real argument greater than 1. 
Appendix B discusses the evaluation of such functions.
\chapter{Evaluating The Infinite Integral Over Four Spherical Bessel Functions}
The integral
$$\int_0^\infty \,r^2\,j_\Llo(k_1r)\,j_\Llt(k_2r)\,
j_\Llth(k_3r)\,j_\Llf(k_4r)\,dr\eqn\TWOPONE$$
where $k_1$, $k_2$, $k_3$ and $k_4$ are positive real numbers,
can be written as
$$\eqalign{&{2\over \pi}\,\int_0^\infty\,K^2\,dK\,\left (
\int_0^\infty\,r^2\,j_\Llo(k_1r)\,j_\Llt(k_2r)\,
j_L(Kr)\,dr\right )
\cr\bspac &\times \left ( \int_0^\infty\,{r^\prime}^2\,
j_\Llth(k_3r^\prime)\,j_\Llf(k_4r^\prime)\,
j_L(Kr^\prime)\,dr \right ),}\eqn\TWOPTWO$$
using the spherical Bessel functions Closure Relation
$$\int_0^\infty\,K^2\,j_L(Kr)\,
j_L(Kr^\prime)\,dK\,=\,{\pi\over 2r^2}\,
\delta (r^\prime \,-\,r).\eqn\TWOPTHREE$$
The value of $L$ is chosen such that it is the smallest value that 
satisfies
$$| \Llo\,-\,\Llt|\,\le\,L\,\le\,\Llo\,+\,\Llt, \eqn\TWOPFOUR$$
and
$$|\Llth\,-\,\Llf|\,\le\,L
\,\le\,\Llth\,+\,\Llf.
\eqn\TWOPFIVE$$

Previously [\MEHR-\MEH] it was shown that
$$\eqalign{&\threej{\Llo}{\Llt}{L}
{0}{0}{0}\,\int_0^\infty \,r^2\,j_\Llo (k_1r)\,
j_\Llt (k_2r)\,j_L (Kr)\,dr\,
=\,{\pi\beta(\Delta) \over 4 k_1 k_2 K} \,i^{\Llo+\Llt-L}\cr\bspac
&\times(2L+1)^{1/2}
\left({k_1 \over K}\right)^L \,
\sum_{{\cal L}=0}^L\,{2L \choose 2{\cal L}}^{1/2} 
\left({k_2 \over k_1}\right)^{\cal L} 
\sum_l\, (2l+1)\cr\bspac &\times
\threej{\Llo}{L - {\cal L}}{l}{0}{0}{0}\,
\threej{\Llt}{{\cal L}}{l}{0}{0}{0}\,
\sixj{\Llo}{\Llt}{L}{{\cal L}}{L-{\cal L}}{l}\,
P_l(\Delta),\cr }\eqn\TWOPSIX$$
where
$\Delta \,=\,(k_1^2+k_2^2-K^2)/2k_1k_2$ and 
$\beta (\Delta)=\theta (1-\Delta)\theta (1+\Delta)$ with $\theta$ the Heaviside function in half-maximum convention. $P_l(x)$ is a Legendre
polynomial of degree $l$, $\threej{\Llo}{\Llt}{L}
{0}{0}{0}$ is a 3j symbol and $\sixj{\Llo}{\Llt}{L}{{\cal L}}
{L-{\cal L}}{l}$ is a 6j symbol which can be found in any standard angular momentum text [\BRNKSA-\EDMNDS]. 
Similarly
$$\eqalign{&\threej{\Llth}{\Llf}{L}
{0}{0}{0}\,\int_0^\infty \,r^2\,j_\Llth (k_3r)\,j_\Llf (k_4r)\,
j_L (Kr)\,dr\,
=\,{\pi\beta(\Delta^\prime)\over 4 k_3 k_4 K} \,
i^{\Llth+\Llf-L}\cr\bspac &\times(2L+1)^{1/2}
\left({k_3 \over K}\right)^{L} \,\sum_{{\cal L}^\prime=0}^
{L}\,
{2L \choose 2{\cal L}^\prime}^{1/2} 
\left({k_4 \over k_3}\right)^{{\cal L}^\prime} 
\sum_{l^\prime}\, (2l^\prime+1)\cr\bspac &\times
\threej{\Llth}{L  - {\cal L}^\prime}{l^\prime}{0}{0}{0}
\,\threej{\Llf}{{\cal L}^\prime}{l^\prime}{0}{0}{0}\,
\sixj{\Llth}{\Llf}{L}{{\cal L}^\prime}
{L-{\cal L}^\prime}{l^\prime}\,
P_{l^\prime}(\Delta^\prime),\cr }\eqn\TWOPSEVEN$$

where $\Delta^\prime\,=\,(k_3^2+k_4^2-K^2)/2k_3k_4$.
The result for eq. $\TWOPONE$ is then

$$\eqalign {& \threej {\Llo}{\Llt}{L}{0}{0}{0}\,
\threej {\Llth}{\Llf}{L}{0}{0}{0}\,\int_0^\infty \,
r^{2}\,j_\Llo (k_1r)
\,j_\Llt(k_2r)\,j_\Llth (k_3r)\,j_\Llf (k_4r)\,dr \cr\bspac &=\,
{\pi\,(k_1k_3)^{L -1}\,i^{\Llo\,+\,\Llt\,+\,\Llth\,+\,\Llf\,-2L}
\,\over 8 k_2k_4}\,(2L+1)\,
\sum_{{\cal L}=0}^L\,\sum_{{\cal L}^\prime =0}^{L}\cr\bspac 
&\times
{2L \choose 2{\cal L}}^{1/2}\,
{2L \choose 2{\cal L}^\prime}^{1/2}
\,(k_2/k_1)^{\cal L}\,(k_4/k_3)^{{\cal L}^\prime}\, 
\sum_l\,\sum_{l^\prime}\,(2l+1)\,(2l^\prime +1)\cr\bspac &\times
\threej {\Llo}{L-{\cal L}}{l}{0}{0}{0}\,
\threej {\Llth}{L-{\cal L}^\prime}{l^\prime}{0}{0}{0}
\,\threej {\Llt}{{\cal L}}{l}{0}{0}{0}\,
\threej {\Llf}{{\cal L}^\prime}{l^\prime}{0}{0}{0}\cr\bspac &\times
\sixj {\Llo}{\Llt}{L}{{\cal L}}{L-{\cal L}}{l}\,
\sixj {\Llth}{\Llf}{L}{{\cal L}^\prime}
{L-{\cal L}^\prime}{l^\prime}\,
J(k_1,k_2,k_3,k_4;l,l^\prime,L).}
\eqn\TWOPEIGHT$$

where
$$J(k_1,k_2,k_3,k_4;l,l^\prime,L)\,\equiv \,\int_0^\infty\,{\beta(\Delta)\,\beta(\Delta^\prime)\over 
K^{2L}}
\,P_l(\Delta)\,P_{l^\prime}(\Delta^\prime)\,dK.\eqn\TWOPNINE$$
This integral can be analytically evaluated for the special case when
$k_3\,=\,k_1$ and $k_4\,=\,k_2$, i.e. $\Delta^\prime\,=\,\Delta$ 
as follows:

Transforming the variable of integration in $\TWOPNINE$ from $K$ to $\Delta$ the integral becomes
$$J(k_1,k_2,k_1,k_2;l,l^\prime,L)\,=\,{k_1k_2\over (2k_1k_2)^{L+1/2}}\,\int_{-1}^1\,
{P_l(\Delta)\,
P_{l^\prime}(\Delta)\over (y\,-\,\Delta)^{L+1/2}}\,
d\Delta,\eqn\TWOPTEN$$
where $y\,\equiv\,(k_1^2\,+\,k_2^2)/2k_1k_2$.

Using appendix A, $\TWOPTEN$ reduces to
$$\eqalign{&J(k_1,k_2,k_1,k_2;l,l^\prime,L)\,=\,
{\sqrt{\pi}\over 
\Gamma (L+1/2)}\,
{\sqrt{k_1k_2}\over |k_1^2\,-\,k_2^2|^{L }}\cr\bspac &\times
\sum_\mu \,(2\mu\,+\,1)\, \Gamma (L+\mu+1/2)\, 
{\threej {l}{l^\prime}{\mu}{0}{0}{0}}^2
\,P_{L}^{-\mu-1/2}
\left ({k_1^2\,+\,k_2^2\over |k_1^2\,-\,k_2^2|}\right ).}
\eqn\TWOPELEVEN$$
The associated Legendre function, $P_{L}^{-\mu-1/2}
\left ({k_1^2\,+\,k_2^2\over |k_1^2\,-\,k_2^2|}\right )$, has 
an integer degree and a half-integer order with a real argument that
is greater than 1. Appendix B discusses the associated Legendre function for such cases and shows closed form expressions for special
cases of $L$.
Furthermore, if $\Llt\,=\,\Llo\equiv l_1$, 
$\Llf\,=\,\Llth \equiv l_2$, the ideal choice 
for $L$ is $L\,=\,0$. Eq. $\TWOPONE$ then becomes
$$\int_0^\infty \,r^2\,j_{l_1}(k_1r)\,j_{l_1}(k_2r)\,j_{l_2}(k_1r)\,
j_{l_2}(k_2r)\,dr\,=\,{\pi\over 4}\,\sum_\mu\,
{\threej {l_1}{l_2}{\mu}{0}{0}{0}}^2\,{k_<^{\mu -1}\over 
k_>^{\mu+2}},\eqn\TWOPTWELVE$$
where $k_<$ ($k_>$) is the smaller (larger) of $k_1$ and $k_2$.
\chapter{Conclusions}
The integral over four spherical Bessel functions 
$$\int_0^\infty \,r^{2}\,j_\Llo(k_1r)\,j_\Llt(k_2r)\,
j_\Llth(k_1r)\,j_\Llf(k_2r)\,dr\eqn\TWOPONE$$
was analytically evaluated for positive integer $\Llo$, $\Llt$, 
$\Llth$ and $\Llf$. The resulting expression involved
a finite sum over an associated Legendre function with integer
degree and half-integer order. An interesting compact result 
was found upon setting $\Llt\,=\,\Llo$ and $\Llf\,=\,\Llth$.

\endpage
\APPENDIX{A.}{A: Evaluation of the Integral $\int_{-1}^1\,
{P_{l}(\Delta)\,
P_{l^\prime}(\Delta)\over (y\,-\,\Delta)^{L+1/2}}\,d\Delta$}

Starting with the integral from [\GRADST], eq. 7.228, page 830
$$\int_{-1}^1 \,{P_\mu (x)\over (y-x)^{L+1/2}}\,dx
\,=\,{2\over \Gamma(L+1/2)}\,(y^2\,-\,1)^{-(L-1/2)/2}\,
e^{-i\pi(L-1/2)}\,Q_\mu^{L-1/2}(y),\eqn\APONE$$
where $Q_\mu^{L-1/2}(y)$ is a Legendre function of the 
second kind of degree $\mu$ and order $L-1/2$. 
Also, using [\GRADST], eq. 8.739, page 1023 gives
$$e^{-i\pi (L-1/2)}\,Q_\mu^{L-1/2}(y)\,=\,\sqrt{{\pi\over 2}}\,
{\Gamma (L\,+\,\mu \, +\,1/2)\over (y^2\,-\,1)^{1/4}}\,
P_L^{-\mu-1/2}({y\over \sqrt{y^2\,-\,1}}).\eqn\APTWO$$

Hence, the itegral in $\APONE$ becomes
$$\int_{-1}^1 \,{P_\mu (x)\over (y-x)^{L+1/2}}\,dx\,=
\,\sqrt{2\pi}\,{\Gamma (L+\mu+1/2)\over \Gamma (L+1/2)}\,
(y^2\,-\,1)^{-L/2}\,P_L^{-\mu-1/2}
({y\over \sqrt{y^2\,-\,1}}).\eqn\APTHREE$$

Multiplying eq. $\APTHREE$ by $(2\mu\,+\,1)\,P_\mu(\Delta)$ 
and summing over $\mu$ results in
$$\eqalign{&{1\over (y\,-\,\Delta)^{L+1/2}}\,=\,\sqrt{{\pi\over 2}}\,
{(y^2\,-\,1)^{-L/2}\over \Gamma (L+1/2)}\,\sum_\mu\,(2\mu\,+\,1)\,
\Gamma (L+\mu+1/2)\,P_\mu(\Delta)\cr\bspac &\times P_L^{-\mu-1/2}
({y\over \sqrt{y^2\,-\,1}}),}\eqn\APFOUR$$
using
$$\sum_\mu \,(2\mu\,+\,1)\,P_\mu(\Delta)\,P_\mu(x)\,=\,2\,
\delta (\Delta\,-\,x).\eqn\APFIVE$$
Hence
$$\eqalign{&{P_l(\Delta)\over(y\,-\,\Delta)^{L+1/2}}\,=\,
\sqrt{{\pi\over 2}}\,
{(y^2\,-\,1)^{-L/2}\over \Gamma (L+1/2)}\,\sum_\mu\,(2\mu\,+\,1)\,
\Gamma (L+\mu+1/2)\cr\bspac &\times
P_L^{-\mu-1/2}({y\over \sqrt{y^2\,-\,1}})\,
\sum_{L^\prime}\,(2L^\prime\,+\,1)\,
{\threej {\mu}{l}{L^\prime}{0}{0}{0}}^2\,
P_{L^\prime}(\Delta),}\eqn\APSIX$$
using
$$\eqalign{&P_\mu^m(\Delta)\,P_\nu^M(\Delta)\,=\,(-1)^{m+M}\,
\sqrt{{(\mu+m)!\,(\nu+M)!\over (\mu-m)!\,(\nu-M)!}}\cr\bspac &\times
\sum_{L^\prime}\,(2L^\prime\,+\,1)\,
\sqrt{{(L^\prime -m-M)!\over (L^\prime +m+M)!}} 
\,\threej {\mu}{\nu}{L^\prime}{0}{0}{0}\,
\threej {\mu}{\nu}{L^\prime}{m}{M}{-m-M}\,
P_{L^\prime}^{m+M}(\Delta).}\eqn\APSEVEN$$

So, multiplying eq. $\APSIX$ by $P_{l^\prime}(\Delta)$ 
and integrating over $\Delta$ from $-1$ to $1$ results in
$$\eqalign{&\int_{-1}^1\,{
P_l(\Delta)\,P_{l^\prime}(\Delta)
\over(y\,-\,\Delta)^{L+1/2}}\,d\Delta\,=\,
\sqrt{2\pi}\,
{(y^2\,-\,1)^{-L/2}\over \Gamma(L+1/2)}
\,\sum_\mu \,(2\mu\,+\,1) \cr\bspac &\times
\Gamma (L+\mu+1/2)\,{\threej {l}{l^\prime}{\mu}{0}{0}{0}}^2 \,
P_L^{-\mu-1/2}({y\over \sqrt {y^2\,-\,1}}),}\eqn\APEIGHT$$
using

$$\int_{-1}^1\,P_{l^\prime}(\Delta)\,P_{L^\prime}(\Delta)\,d\Delta
\,=\,{2\over 2l^\prime +1}\,\delta_{L^\prime,~l^\prime}.\eqn\APNINE$$
\endpage
\APPENDIX{B.}{B: Associated Legendre Fumctions, $P_l^m(x)$ for $x>1$}
Previously we have shown that associated Legendre functions, 
$P_l^m(x)$, for integer $l$ and real $m$ can be written as
$$P_l^m(x)={(-1)^l\over 2^l \Gamma (|l-m|+1)}\,
({1-x \over 1+x})^{m/2}\,{d^l \over dx^l}\,[(1-x^2)^l({1+x\over 1-x})^m],\eqn\BPONE$$
for real $-1<x<1$, and the factorial is replaced by 
the gamma function 
to allow for non-integer $m$. The special cases for particular $l$ values were
$$P_0^m(x)\,=\,{1\over \Gamma (|m|+1)}\,\,({1+x\over 1-x})^{m/2},
\eqn\BPTWO$$
$$P_1^m(x)\,=\,\,{1\over \Gamma (|1-m|+1)}\,\,(x-m)\,\,
({1+x\over1-x})^{m/2},\eqn\BPTHREE$$
$$P_2^m(x)\,=\,{1\over \Gamma (|2-m|+1)}\,\,(3\,{x}^{2}-3\,xm-1+{m}^{2})\,({1+x\over 1-x})^{m/2},\eqn\BPFOUR$$
$$P_3^m(x)\,=\,{1\over \Gamma (|3-m|+1)}\,\,(15\,{x}^{3}-15\,
{x}^{2}m-9\,x+6\,x{m}^{2}+4\,m-{m}^{3})\,({1+x\over 1-x})^{m/2},
\eqn\BPFIVE$$
$$\eqalign{&P_4^m(x)\,=\,{1\over \Gamma (|4-m|+1)}\,\,(105\,
{x}^{4}-105\,{x}^{3}m-90\,{x}^{2}+45\,{x}^{2}{m}^{2}+55\,xm\cr
\bspac
&-10x{m}^{3}+9-10\,{m}^{2}+{m}^{4})\,({1+x\over 1-x})^{m/2},}
\eqn\BPSIX$$
Now, by comparing equations 8.702, page 1014 and 8.704, page 1015 of [\GRADST], one finds that for real $x>1$
$$P_l^m(x)={(-1)^l\over 2^l \Gamma (|l-m|+1)}\,
({x-1 \over x+1})^{m/2}\,{d^l \over dx^l}\,[(1-x^2)^l({x+1\over x-1})^m],\eqn\BPSEVEN$$
with the special cases
$$P_0^m(x)\,=\,{1\over \Gamma (|m|+1)}\,\,({x+1\over x-1})^{m/2},
\eqn\BPEIGHT$$
$$P_1^m(x)\,=\,\,{1\over \Gamma (|1-m|+1)}\,\,(x-m)\,\,
({x+1\over x-1})^{m/2},\eqn\BPNINE$$
$$P_2^m(x)\,=\,{1\over \Gamma (|2-m|+1)}\,\,(3\,{x}^{2}-3\,xm-1+{m}^{2})\,({x+1\over x-1})^{m/2},\eqn\BPTEN$$
$$P_3^m(x)\,=\,{1\over \Gamma (|3-m|+1)}\,\,(15\,{x}^{3}-15\,
{x}^{2}m-9\,x+6\,x{m}^{2}+4\,m-{m}^{3})\,({x+1\over x-1})^{m/2},
\eqn\BPELEVEN$$
$$\eqalign{&P_4^m(x)\,=\,{1\over \Gamma (|4-m|+1)}\,\,(105\,
{x}^{4}-105\,{x}^{3}m-90\,{x}^{2}+45\,{x}^{2}{m}^{2}+55\,xm\cr
\bspac
&-10x{m}^{3}+9-10\,{m}^{2}+{m}^{4})\,({x+1\over x-1})^{m/2},}
\eqn\BPTWELVE$$
\endpage
\par \penalty-400 \vskip\chapterskip
   \spacecheck\referenceminspace \immediate\closeout\referencewrite
   \referenceopenfalse
   \line{\fourteenrm\hfil References\hfil}\vskip\headskip
   \input referenc.tex

%
\bye